\documentclass[aps,prd,twocolumn,groupedaddress,amssymb,eqsecnum,showpacs,epsfig,nofootinbib]{revtex4}
\usepackage{graphicx}
\usepackage{bm}
\usepackage{dcolumn}
\usepackage{amsmath}

\usepackage{amsmath}
\usepackage{amssymb}

\numberwithin{equation}{section}
\usepackage{epsfig}

\def \lleq {\lower0.9ex\hbox{ $\buildrel < \over \sim$} ~}
\def \ggeq {\lower0.9ex\hbox{ $\buildrel > \over \sim$} ~}

\def \beq  {\begin{equation}}
\def \eeq  {\end{equation}}
\def \ber  {\begin{eqnarray}}
\def \eer  {\end{eqnarray}}

\begin{document}
\newcommand{\newc}{\newcommand}

\newc{\be}{\begin{equation}}
\newc{\ee}{\end{equation}}
\newc{\ba}{\begin{eqnarray}}
\newc{\ea}{\end{eqnarray}}
\newc{\bea}{\begin{eqnarray*}}
\newc{\eea}{\end{eqnarray*}}
\newc{\D}{\partial}
\newc{\ie}{{\it i.e.} }
\newc{\eg}{{\it e.g.} }
\newc{\etc}{{\it etc.} }
\newc{\etal}{{\it et al.}}
\newc{\lcdm }{$\Lambda$CDM }
\newcommand{\nn}{\nonumber}
\newc{\ra}{\rightarrow}
\newc{\lra}{\leftrightarrow}
\newc{\lsim}{\buildrel{<}\over{\sim}}
\newc{\gsim}{\buildrel{>}\over{\sim}}
\title{Kink-Antikink Formation from an Oscillation Mode by Sudden Distortion of the Evolution Potential}
\author{C. S. Carvalho$^a$ and L. Perivolaropoulos$^b$}
%\email{leandros@uoi.gr}
 \affiliation{$^a$Astrophysics and Cosmology Research
Unit, School of Mathematical Sciences, University of KwaZulu-Natal, Durban, 4041, South Africa.\\$^b$Department of Physics, University of Ioannina, Greece.}
\date{\today}

\begin{abstract}
We demonstrate numerically that an oscillation mode in 1+1 dimensions (eg a breather or an oscillon) can decay into a kink-antikink pair by a sudden distortion of the evolution potential which occurs within a certain time or space domain. In particular, we consider the transition of a sine-Gordon potential into a $\Phi^4$ potential. The breather field configuration is assumed to initially evolve in a sine-Gordon potential with velocity $v$ and oscillation frequency $\omega$. We then consider two types of numerical experiments: a. An abrupt transition of the potential to a $\Phi^4$ form at $t_0=0$ over the whole 1-dimensional lattice and b. The impact of the breather on a region $x>x_0=0$ where the potential has the $\Phi^4$ form which is different from the sine-Gordon form valid at $x<x_0=0$. We find that in both cases there is a region of parameters $(v,\omega)$ such that the breather decays to a kink-antikink pair. This region of parameters for kink-antikink formation is qualitatively similar with the parameter region where the energy of the breather exceeds the energy of the kink-antikink pair in the $\Phi^4$ potential. We demonstrate that the same mechanism for soliton formation is realized when using a gaussian oscillator (oscillon) instead of a breather. We briefly discuss the implications of our results for realistic experiments as well as their extension to soliton formation in two and three space dimensions.
\end{abstract}
%
%\pacs{98.80.Es,98.65.Dx,98.62.Sb}
%
\maketitle

\section{Introduction}

Topological solitons (defects) \cite{Rajaramanbook,Vachaspatibook,Tong:2005un} are static localized solutions of nonlinear partial differential equations which are stable due to nontrivial topological properties of the vacuum manifold. These coherent non-perturbative excitations are distinct from the perturbative `particle' excitations which correspond to localized small field oscillations around the vacuum \cite{Dutta:2008jt}.

Solitons arise in the context of effective field theories \cite{Bunkov:2000,Arodz:2003} (eg Ginzburg-Landau theory) and play an important role in many branches of modern physics. For example they form in condensed matter systems\cite{def-condmat} (liquid Helium, liquid crystals or superconductors) and in cosmology \cite{Hindmarsh:1994re,Durrer:2001cg,Perivolaropoulos:2005wa,Magueijo:2000se} (cosmic strings, domain walls, monopoles) in the context of high energy physics (HEP) models. Solitons usually form during phase transitions in condensed matter or cosmological systems. In the context of HEP models, these phase transitions are only realized in the early universe and it is currently not possible to reproduce them in the laboratory or in accelerators. Thus, the experimental  study of the formation of topological defects predicted in HEP models can only be made by drawing analogies with condensed matter systems where the experimental realization of such phase transitions is possible \cite{Bunkov:2000,def-condmat}.

The understanding of the transition between the `particle' and the `soliton' sectors is an important unsolved problem. The development of methods  that would allow the formation of solitons from particles either through scattering or through decay would open new possibilities for the experimental study of solitons predicted by HEP theories.

Previous approaches to this problem have focused on the formation of solitons from scattering of two \cite{Levkov:2004tf,Mattis:1991bj} or more \cite{Dutta:2008jt} particles. These approaches are motivated by a time reversal of a soliton-antisoliton annihilation which is accompanied by emission of oscillation modes (particles). In the case of two particle scattering, it has been shown that the soliton formation is exponentially suppressed \cite{Levkov:2004tf} even though it may proceed more efficiently in the presence of a pre-existing soliton \cite{Manton:1996ex,Romanczukiewicz:2005rm}. A multi-particle scattering has been shown to be more effective in producing solitons \cite{Dutta:2008jt} through resonant build-up of particle oscillations which allows the oscillations to extend to other vacua. The initial conditions for this process however require careful tuning in order to achieve the required resonance \cite{Dutta:2008jt}.
\begin{figure*}[ht]
\centering
\begin{center}
$\begin{array}{@{\hspace{-0.10in}}c@{\hspace{0.0in}}c}
\multicolumn{1}{l}{\mbox{}} &
\multicolumn{1}{l}{\mbox{}} \\ [-0.2in]
\epsfxsize=3.3in
\epsffile{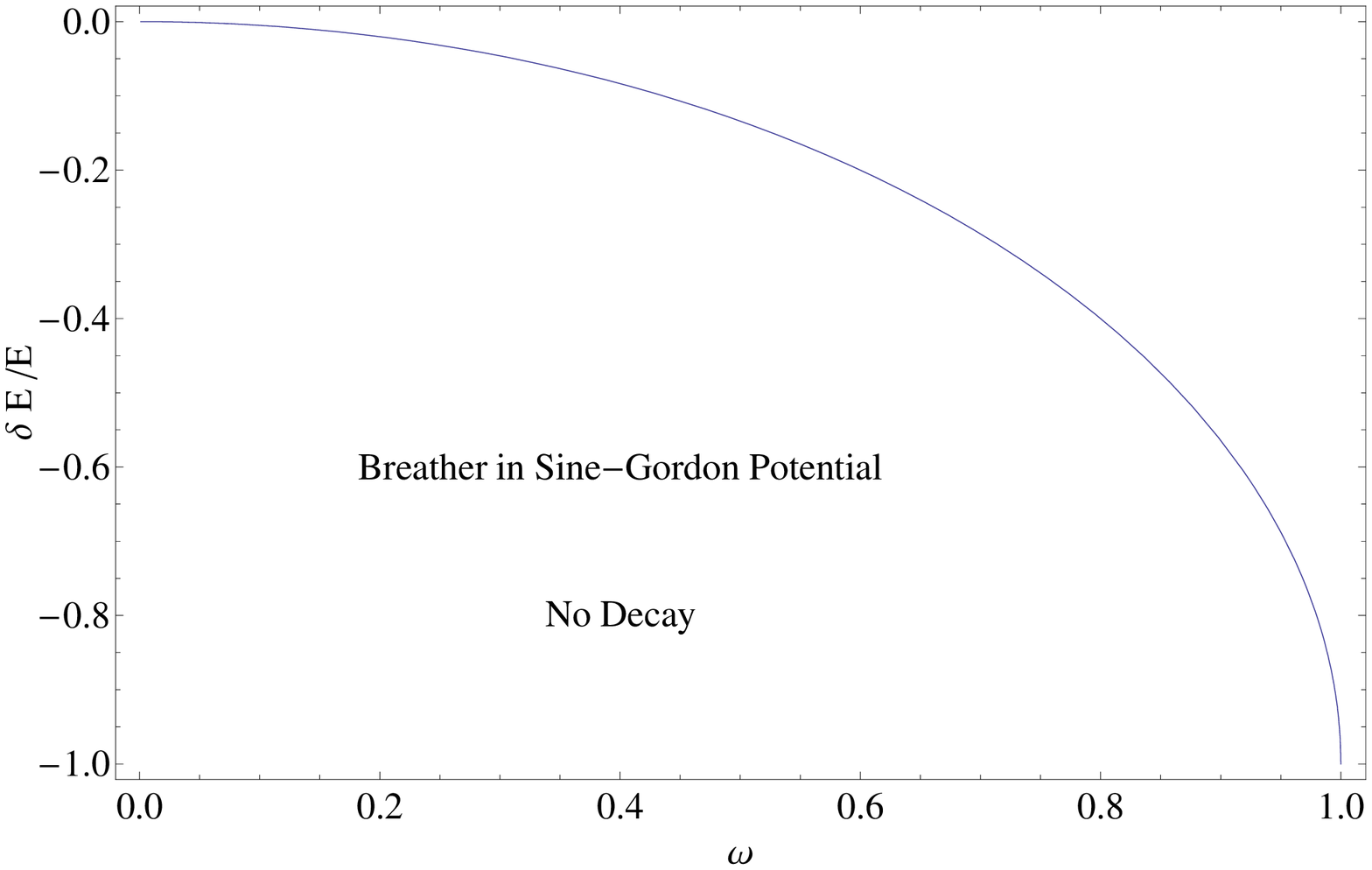} &
\epsfxsize=3.3in
\epsffile{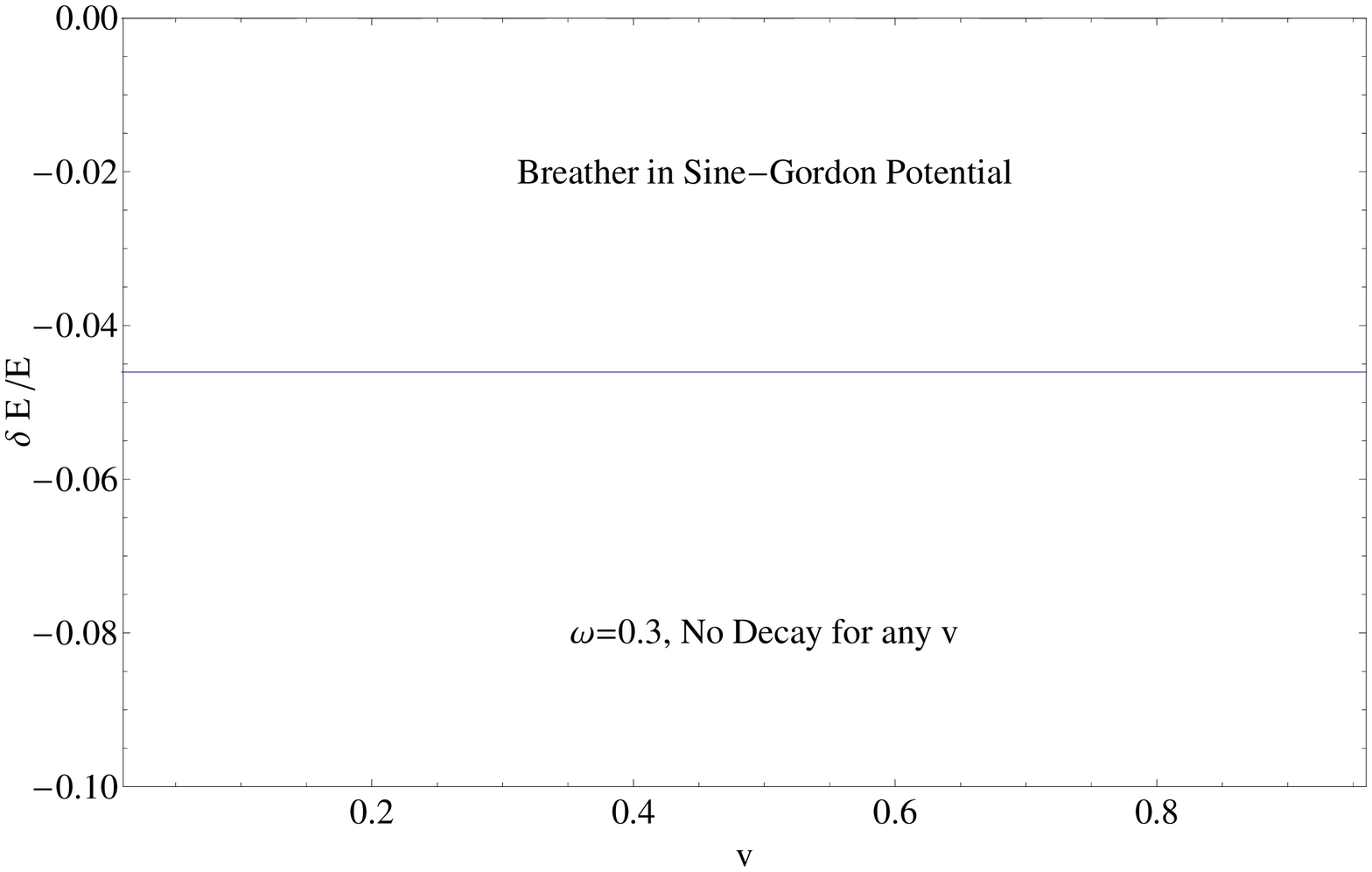} \\
\end{array}$
\end{center}
\vspace{0.0cm}
\caption{\small a: The energy of the breather in the sine-Gordon Lagrangian is smaller than the energy of a kink-antikink pair for all values of $\omega$ (see also eq. (\ref{dedef})). Here we used $v=0.3$ but a similar plot is obtained for all values of $v$. b: The relative energy difference of the breather in the sine-Gordon Lagrangian with respect to the sine-Gordon kink-antikink pair is independent of the boost velocity since both field configurations (breather and kinks) are exact solutions of the sine-Gordon Lagrangian.}
\label{fig1}
\end{figure*}

An alternative approach to the problem of soliton formation involves the decay of a single highly energetic oscillation mode (particle) to a soliton-antisoliton pair (eg kink-antikink in 1+1 dimensions). This decay may be facilitated by a time or space dependent distortion of the potential determining the dynamical evolution of the oscillation mode. Such a distortion may be achieved either by changing an external parameter of a system (pressure, temperature, external field etc) or, in the space dependent case, by considering two different materials separated by a given surface.

The type of distortion considered in the present study is distinct from the corresponding distortion occurring in the effective potential during a phase transition. First, our initial conditions are not thermal (they consist of an oscillator mode) and second, we only consider mild forms of distortions that do not change the topological properties of the vacuum. In particular, we consider the transition from a sine-Gordon potential to a $\Phi^4$ potential in $1+1$ dimensions occuring at a given time $t=t_0$ or a given spatial point $x=x_0$. The initial field configuration is an oscillation mode in the form of either a sine-Gordon breather or a gaussian oscillator (`oscillon'). The energy of the initial field configuration as recorded by the distorted potential is tuned by the boost velocity, the field configuration parameters (eg the oscillation frequency) and the phase of the configuration at the space-time point of the transition (ie the value of $t_0$ or of $x_0$).

An important condition required for the decay of the initial oscillation mode to a kink-antikink pair is the fact that the energy of the oscillation mode as recorded by the distorted potential ($\Phi^4$) should be larger than the energy of the kink-antikink pair in the same potential. This necessary conditions is of the form
\be \frac{\delta E}{E} (v,\omega,t_0) \equiv \frac{E_{om}^{\Phi^4} (v,\omega,t_0)}{2 E_{kink}^{\Phi^4}(v=0)}-1\geq 0 \label{decond} \ee
where $E_{om}^{\Phi^4}$ is the energy of the boosted oscillation mode evaluated in the distorted potential ($\Phi^4$) at the transition time $t_0$ and $E_{kink}^{\Phi^4}(v=0)$ is the corresponding energy of the kinks in the same potential. We have considered the conservative assumption that the kink-antikink pair forms at 0 velocity. In practice, momentum conservation implies that the kink-antikink velocity is non zero. This is expected to decrease somewhat the allowed parameter region for soliton formation.

The goal of the present paper is to investigate the decay of a particle oscillation mode into kink-antikink pairs in the context of a $\Phi^4$ potential in $1+1$ dimensions. We investigate the $v-\omega$ (boost velocity-oscillation frequency) parameter region where the decay occurs and consider two types of particle-like oscillators \begin{itemize} \item A sine-Gordon breather \item A gaussian oscillator (oscillon) \end{itemize} We compare the theoretically allowed (based on energetics) parameter region of decay with the corresponding region based on numerical simulations of field evolution and we find qualitative agreement. We also demonstrate numerically that the decay of the boosted sine-Gordon breather in the context of the sine-Gordon Lagrangian (without switch to $\Phi^4$) does not occur for any boost velocity as expected by energetic arguments and Lotentz invariance since \be \frac{\delta E}{E} (v,\omega,t_0) \equiv \frac{E_{om}^{sG} (v,\omega)}{2 E_{kink}^{sG}(v)}-1< 0 \label{decond1} \ee for any boost velocity $v$.

\section{Time Dependent Distortions of the Potential}

As discussed in the Introduction, particles at the classical level may be represented as small field oscillations around the vacuum. At the quantum level, these oscillations are quantized and the energy of the lowest quantum state is equal to that of a particle. Thus, the lowest energy oscillation mode may be identified with the particle excitation of the theory. A useful field configuration that can play the role of an oscillation mode at the classical level is the breather solution of the sine-Gordon model\cite{Rajaramanbook}. The sine-Gordon Lagrangian is of the form
\be
 L _{sG} =  \frac{1}{2} \partial_\mu \Phi \partial^\mu \Phi -
           \frac{1}{\pi^2} [ 1+\cos (\pi \Phi)]
 \label{sglang} \ee with field equation
 \be
{\ddot \Phi} = \Phi'' + \frac{1}{\pi}\sin (\pi   \Phi)
\label{eom}
\ee
The breather solution is given by \cite{Rajaramanbook}
 \be
\Phi^{sG}_b (t,x; \omega , v) = -1+
    \frac{4}{\pi} \tan^{-1} \left [ \frac{\eta \sin(\omega T)}
                            {\cosh (\eta \omega X)} \right ]
\label{br-sol}
\ee
where
\begin{eqnarray}
T &=& \gamma [t - v (x-x_0)] \ , \ \ X = \gamma [x-x_0 - v t]
\nonumber \\
\gamma &=& (1-v^2)^{-1/2} \ ,  \ \ \eta = \sqrt{1-\omega^2} /\omega  \, .
\end{eqnarray}
In addition to the breather, the sine-Gordon Lagrangian has soliton solutions (kinks) that interpolate between neighboring vacua $\Phi_n=2 n +1$, $\Phi_{n+1} = 2n+3$, ($n=...,-1,0,+1,...$). Such a sine-Gordon kink interpolating between the vacua $\Phi_{-1}=-1$ and $\Phi_0 =+1$ is of the form
\be
\phi^{sG}_{kink} (t,x; \omega , v) = -1+
    \frac{4}{\pi} \tan^{-1} \left [ \exp(X) \right ] \label{sg-kink} \ee while the corresponding antikink is ${\bar \Phi}_{kink}^{sG} = - \Phi_{kink}^{sG}$. It is straightforward to show that the sine-Gordon energy is \be
E = \int dx \left [ \frac{\dot\Phi ^2}{2}
                    + \frac{{\Phi '} ^2}{2}
                    + \frac{1}{\pi^2} [ 1+\cos (\pi \Phi)] \right ]
\label{ensg} \ee

\begin{figure}[!t]
\hspace{0pt}\rotatebox{0}{\resizebox{.5\textwidth}{!}{\includegraphics{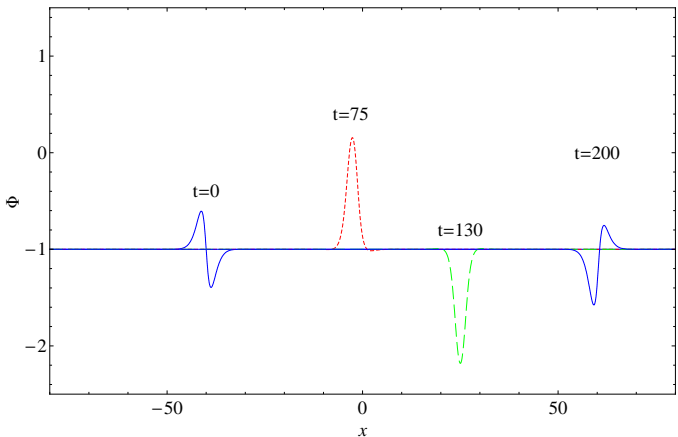}}}
\vspace{0pt}{\caption{The boosted breather evolved by the sine-Gordon Lagrangian for $(v,\omega)=(0.5,0.6)$. As expected, the decay to a kink-antikink pair does not occur for this and for any other parameter value we have tested.}} \label{fig2}
\end{figure}
For the breather (\ref{br-sol}), the energy (\ref{ensg}) is smaller than the corresponding energy of the kink-antikink pair (\ref{sg-kink}) ie
\be \frac{\delta E}{E} (v,\omega) \equiv \frac{E_b^{sG} (v,\omega)}{2 E_{kink}^{sG}(v)}-1=\frac{E_b^{sG} (0,\omega)}{2 E_{kink}^{sG}(0)}-1 \leq 0 \label{dedef} \ee for all values of $v$, $\omega$. This is shown in Fig. 1. We have verified numerically that $\frac{\delta E}{E} (v,\omega)$ is independent of the boost velocity since both the breather and the kink are exact solutions of the sine-Gordon field equation (see Fig. 1b) and therefore $E_b^{sG} (v,\omega)=\gamma E_b(0,\omega)$ and $E_{kink}^{sG} (v)=\gamma E_{kink}(0)$. We have verified the stability of the sine-Gordon breather in the context of the Lagrangian (\ref{sglang}) by numerically simulating its evolution for various values of $v$, $\omega$. Snapshots of such a simulation are shown in Fig. 2.

We now consider the embedding of the breather as an initial condition in a $\Phi^4$ Lagrangian of the form
\be
L = \frac{1}{2} (\partial_\mu \Phi )^2 -
                  \frac{1}{4} \left ( \Phi^2 - 1 \right )^2
\label{f4lagr}
\ee
We will consider a numerical experiment such that the breather evolves in a sine-Gordon Lagrangian (\ref{sglang}) where it is an exact solution for $t<0$ but at $t=t_0=0$ the Lagrangian changes abruptly to (\ref{f4lagr}). Since the evolution for $t<0$ is trivial, we only consider the evolution for $t\geq 0$.
The equation of motion for $t\geq 0$ is
\be
{\ddot \Phi} = \Phi'' - (\Phi^2 -1) \Phi
\label{eomf4}
\ee
and the corresponding energy and momentum are
\be E = \int dx \left [ \frac{\dot\Phi ^2}{2}
                    + \frac{{\Phi '} ^2}{2}
                    + \frac{1}{4} \left ( \Phi^2 - 1 \right )^2 \right ] \label{f4en}
\ee
\be P = -\int dx \dot\Phi \;{\Phi '}
\label{f4mom}
\ee
In addition to perturbative particle excitations around the vacua this equation of motion admits static non-perturbative kink solutions of the form
\be
\Phi_{kink} = {\rm tanh}\left ( \frac{X}{\sqrt{2}} \right )
\ee
which interpolate between the two vacua $\Phi=\pm 1$. One way to mimic the presence of particles in the $\Phi^4$ potential is to use arbitrary field configurations that oscillate around a single vacuum with long lifetimes. One such choice is the sine-Gordon breather (\ref{br-sol}) embedded in the $\Phi^4$ Lagrangian.

\begin{figure}[!t]
\hspace{0pt}\rotatebox{0}{\resizebox{.5\textwidth}{!}{\includegraphics{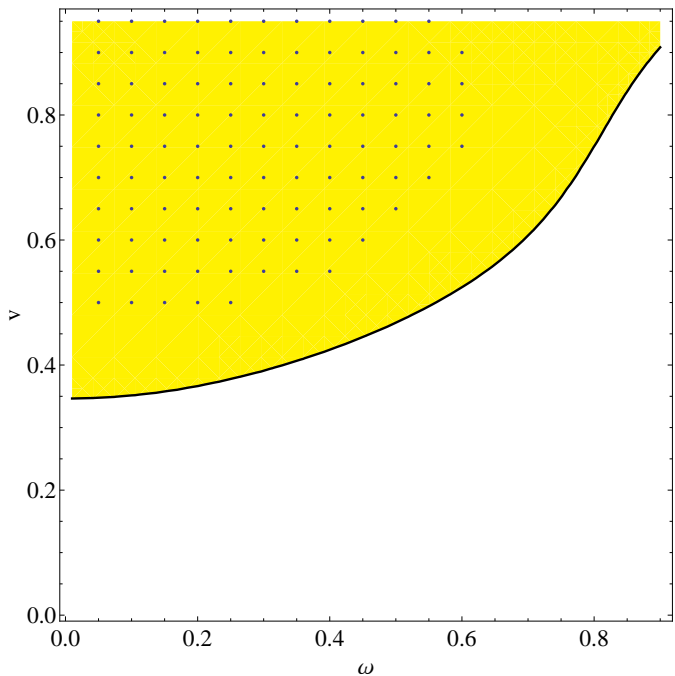}}}
\vspace{0pt}{\caption{The parameter range where $\frac{\delta E}{E}>0$ is indicated by the shaded region. The dots indicate parameter values where the decay of the boosted breather into a kink-antikink pair was observed in the simulations. The whole parameter region shown in the plot was scanned at steps of $\Delta v =0.05$, $\Delta \omega =0.05$.}} \label{fig3}
\end{figure}
%Since the embedded breather 4-momentum does not transform as a 4-vector in contrast with the 4-momentum of the $\Phi^4$ kink,
The sign of $\frac{\delta E}{E}$ depends not only on $v,\omega$ but also on the phase of the breather at the time $t_0$ of the potential switch. Thus $\frac{\delta E}{E}=\frac{\delta E}{E}(v,\omega,t_0)$. In what follows we fix $t_0$ ($t_0=0$) and we focus on the parameters $v,\omega$ demonstrating that kink-antikink formation is possible for a range of these parameters.

The range of boost velocities $v$ and corresponding values of $\omega$ where energetics allow the decay of the boosted breather to a kink-antikink pair is shown in Fig. 3 (shaded region in $v-\omega$ plane).
\begin{figure*}[ht]
\centering
\begin{center}
$\begin{array}{@{\hspace{-0.10in}}c@{\hspace{0.0in}}c}
\multicolumn{1}{l}{\mbox{}} &
\multicolumn{1}{l}{\mbox{}} \\ [-0.2in]
\epsfxsize=3.3in
\epsffile{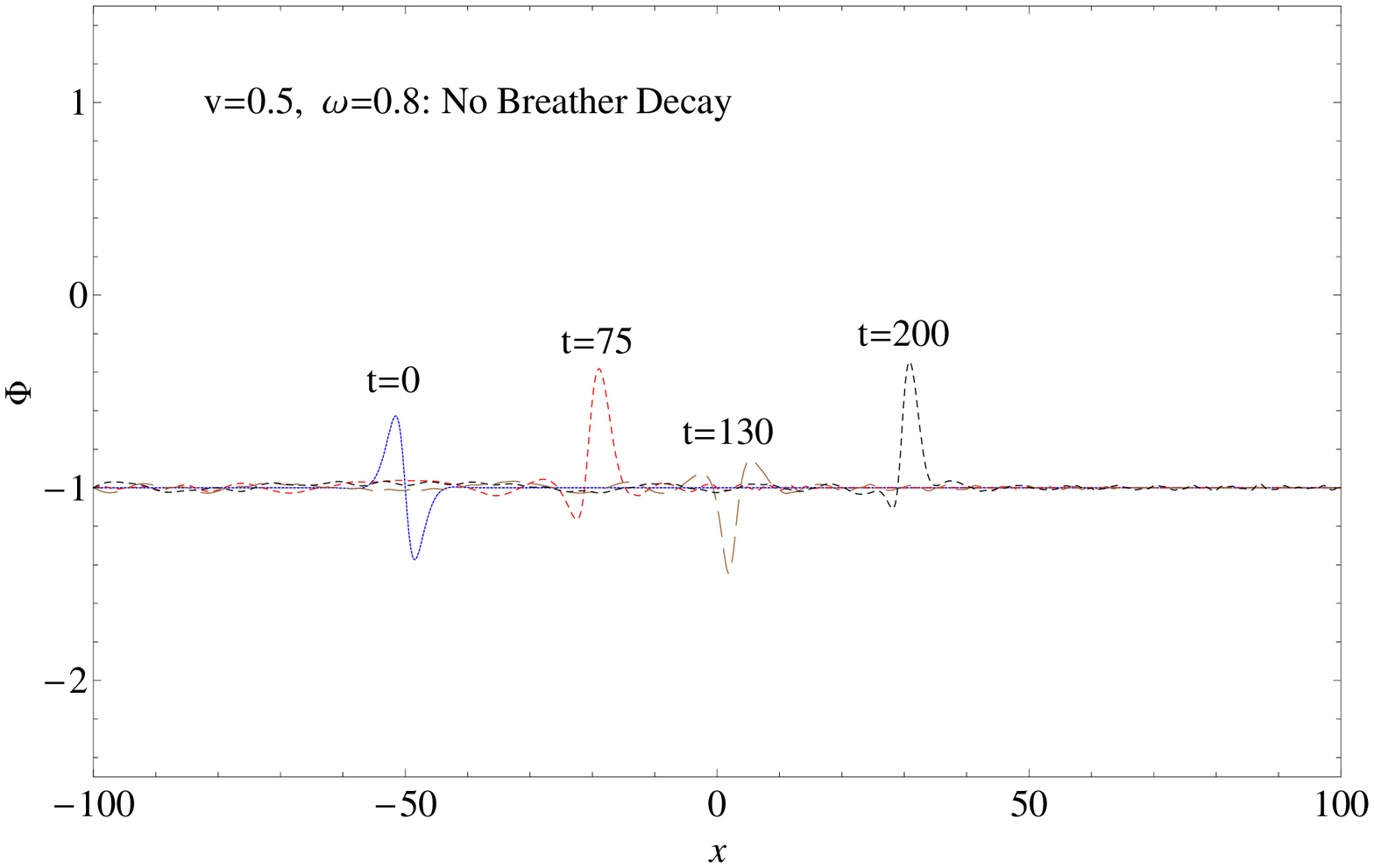} &
\epsfxsize=3.3in
\epsffile{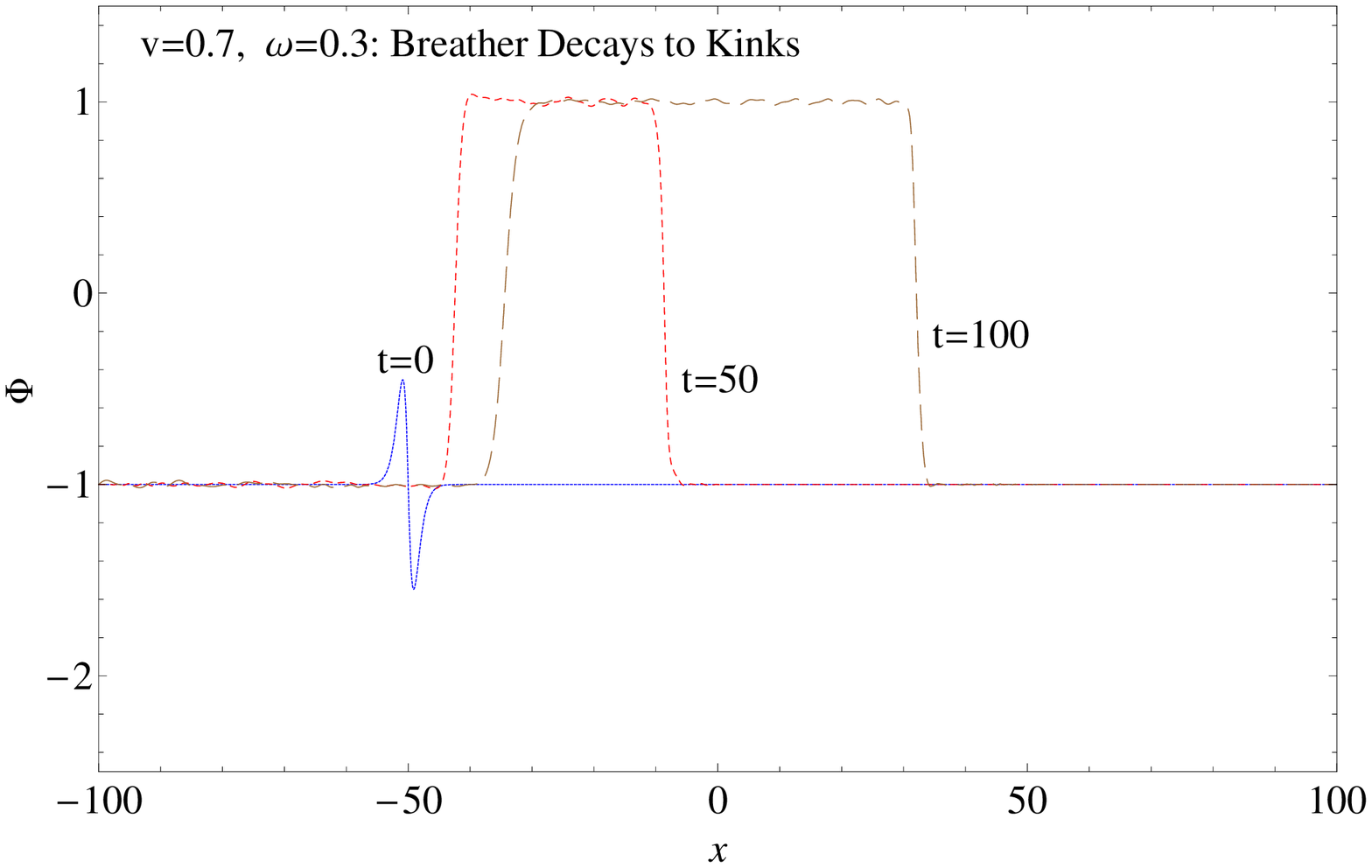} \\
\end{array}$
\end{center}
\vspace{0.0cm}
\caption{\small a: The field evolution of the boosted breather with $(v,\omega)=(0.5,0.8)$, embedded in the $\Phi^4$ Lagrangian. The breather evolves oscillating without decay to solitons since there is not enough energy for the decay. b: The field evolution of the boosted breather with $(v,\omega)=(0.7,0.3)$, embedded in the $\Phi^4$ Lagrangian. In this case, the boost provides sufficient energy for the decay to a kink-antikink pair.}
\label{fig4}
\end{figure*}
Even though, energetics allow the decay of the embedded sine-Gordon breather into a kink-antikink pair for a range of parameters, it is not necessary that this decay will occur in practice. For example the extra energy may be emitted in the form of radiation or remain as breather oscillating energy. In order to find if the decay actually occurs we have performed numerical simulations of the embedded breather evolution spanning the full range of the $v-\omega$ parameters.

We have performed the simulations using the NDSolve routine of Mathematica\cite{mathematica}, solving the partial differential equation (\ref{eomf4}) with the embedded breather (\ref{br-sol}) as initial condition and fixed boundary conditions $\Phi =-1$, ${\dot \Phi}=0$.
Using these simulations we have confirmed that there is a parameter range where the embedded breather spontaneously decays into a kink-antikink pair at the time of the potential switch. In Fig. 4 we show the evolution of the embedded breather field for two parameter sets $v-\omega$ corresponding to stability (Fig. 4a:$(v,\omega)=(0.5,0.8)$) and instability towards kink-antikink decay  (Fig. 4b:$(v,\omega)=(0.7,0.3)$).

We have performed various tests to secure the validity of our numerical simulations. These tests include the following:
\begin{itemize}
\item
We have verified that total energy (eq. (\ref{f4en})) was conserved at a level of about 1\% during the evolution.
\item
We have verified that total momentum (eq. (\ref{f4mom})) was conserved at a level of about 1\% during the evolution.
\item
We have doubled the size of the lattice and verified that the evolved field configurations remain practically unchanged.
\item
We have considered 5 different numerical algorithms for the solution of the partial differential equation (\ref{eomf4}) and they all lead to practically identical field evolution. The algorithms were used through the 'Method' option of the NDSolve routine of Mathematica and include the methods: Adams, ExplicitRungeKutta, ImplicitRungeKutta, DoubleStep with submethod:  ExplicitModifiedMidpoint and Extrapolation with submethod ExplicitEuler .
\end{itemize}
The instability parameter range obtained using simulations is spanned with dots in Fig. 3 and is smaller than the shaded range anticipated from energetic considerations. Indeed, as discussed above, the supply of sufficient energy is a necessary but not a sufficient condition for the kink-antikink formation. For example, center of mass momentum needs to also be conserved for kink-antikink formation while in the energy criterion (\ref{decond}) $\frac{\delta E}{E}$ was defined assuming zero velocity of the pair at formation. The assumption of a nonzero velocity of the kink-antikink at formation would tend to decrease the theoretically allowed parameter region from energetics bringing it to even better agreement with simulations.

The embedded oscillating breather ansatz has the advantage of being an exact solution to a Lagrangian similar to that of the $\Phi^4$ potential which implies a long lifetime and minimal radiation emission during its evolution. On  the other hand, it has a maximum angular frequency $\omega_{max}=1$
and this limits the maximum energy that can be achieved by boosting. As a result we have not observed more complicated decay products than a kink-antikink pair (eg two kink-antikink pairs) since even though we can easily obtain $E_m^{\Phi^4} > 2 M_{kink}$ it is not possible to achieve $E_m^{\Phi^4} > 4 M_{kink}$ which would allow decay to two kink-antikink pairs.

In order to bypass this limitation we have considered an alternative oscillating ansatz corresponding to a gaussian oscillator also known as `oscillon' in previous studies \cite{Bogolyubsky:1976nx,Gleiser:1993pt,Copeland:1995fq,Honda:2000gv,Hindmarsh:2006ur}. This gaussian oscillator field configuration is of the form
\be
\Phi_{go}(t,x;,v,\omega)=-1+a\sin(\omega T) \exp(-b X^2)  \label{gosc} \ee
where $a$, $b$ are parameters that we fix to the values $a=1$ and $b=0.1$ in most of what follows.

\begin{figure}[!t]
\hspace{0pt}\rotatebox{0}{\resizebox{.5\textwidth}{!}{\includegraphics{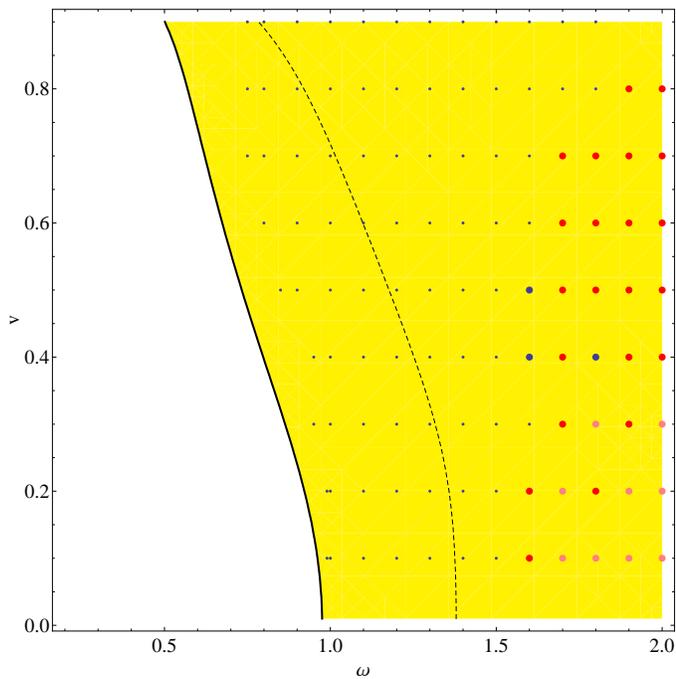}}}
\vspace{0pt}{\caption{The parameter range where $\frac{\delta E}{E}>0$ is indicated by the shaded region. The dots indicate parameter values where the decay of the gaussian oscillator into a kink-antikink pair was observed in the simulations. Thicker dots indicate decay into more complicated products (eg more than one kink-antikink pair or additional particles). The energy of the initial gaussian oscillator is larger than two kink-antikink pairs on the right of the dashed line.}} \label{fig5}
\end{figure}

The $v-\omega$ parameter region where $\frac{\delta E}{E}>0$ and therefore the decay is energetically allowed, is shown in Fig. 5 (shaded region).The region on the right of the dashed line corresponds to parameter values where the  gaussian oscillator energy is larger than two kink-antikink pairs (four solitons) implying the possibility of more complicated decay products.
\begin{figure*}[ht]
\centering
\begin{center}
$\begin{array}{@{\hspace{-0.10in}}c@{\hspace{0.0in}}c}
\multicolumn{1}{l}{\mbox{}} &
\multicolumn{1}{l}{\mbox{}} \\ [-0.2in]
\epsfxsize=3.3in
\epsffile{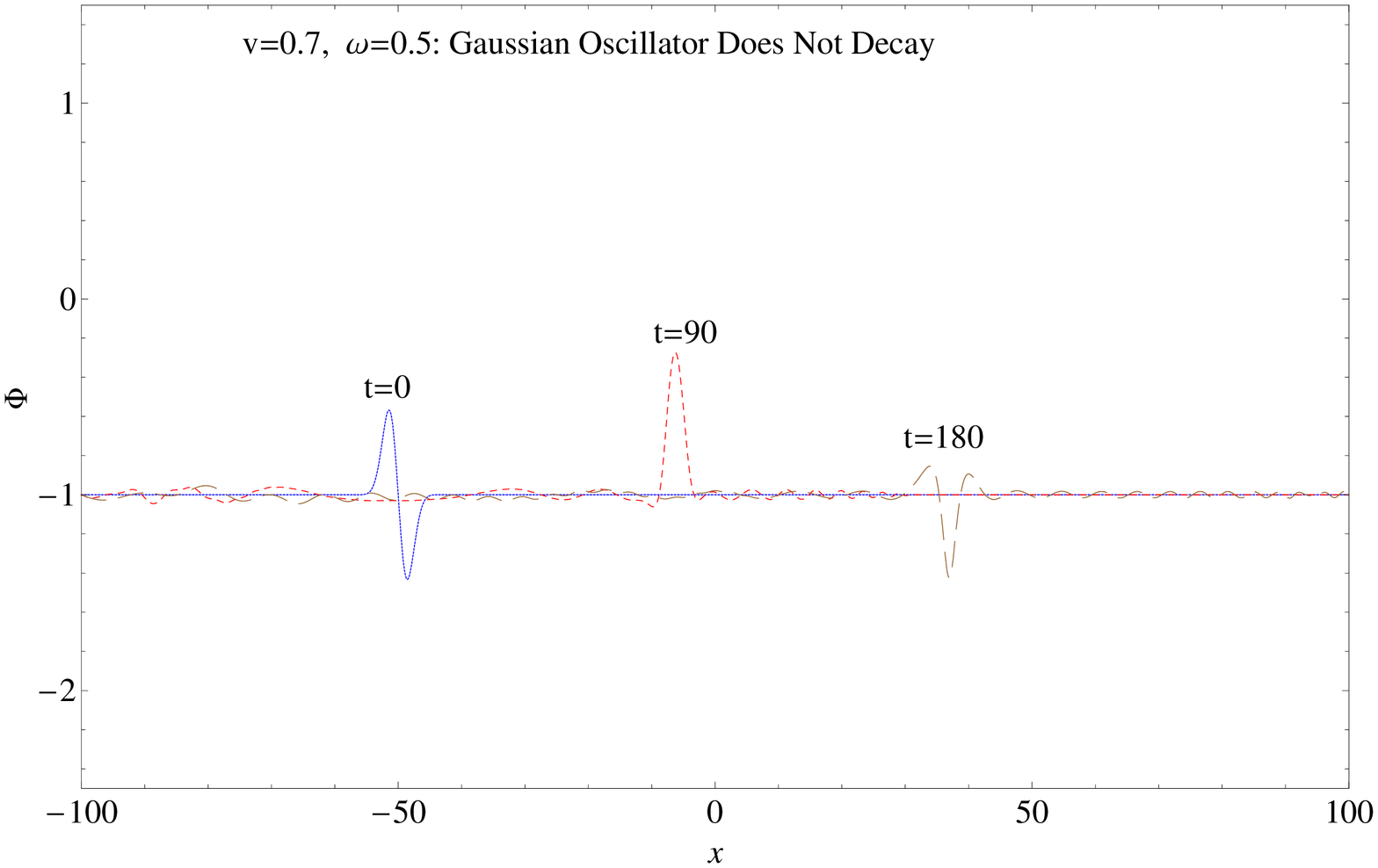} &
\epsfxsize=3.3in
\epsffile{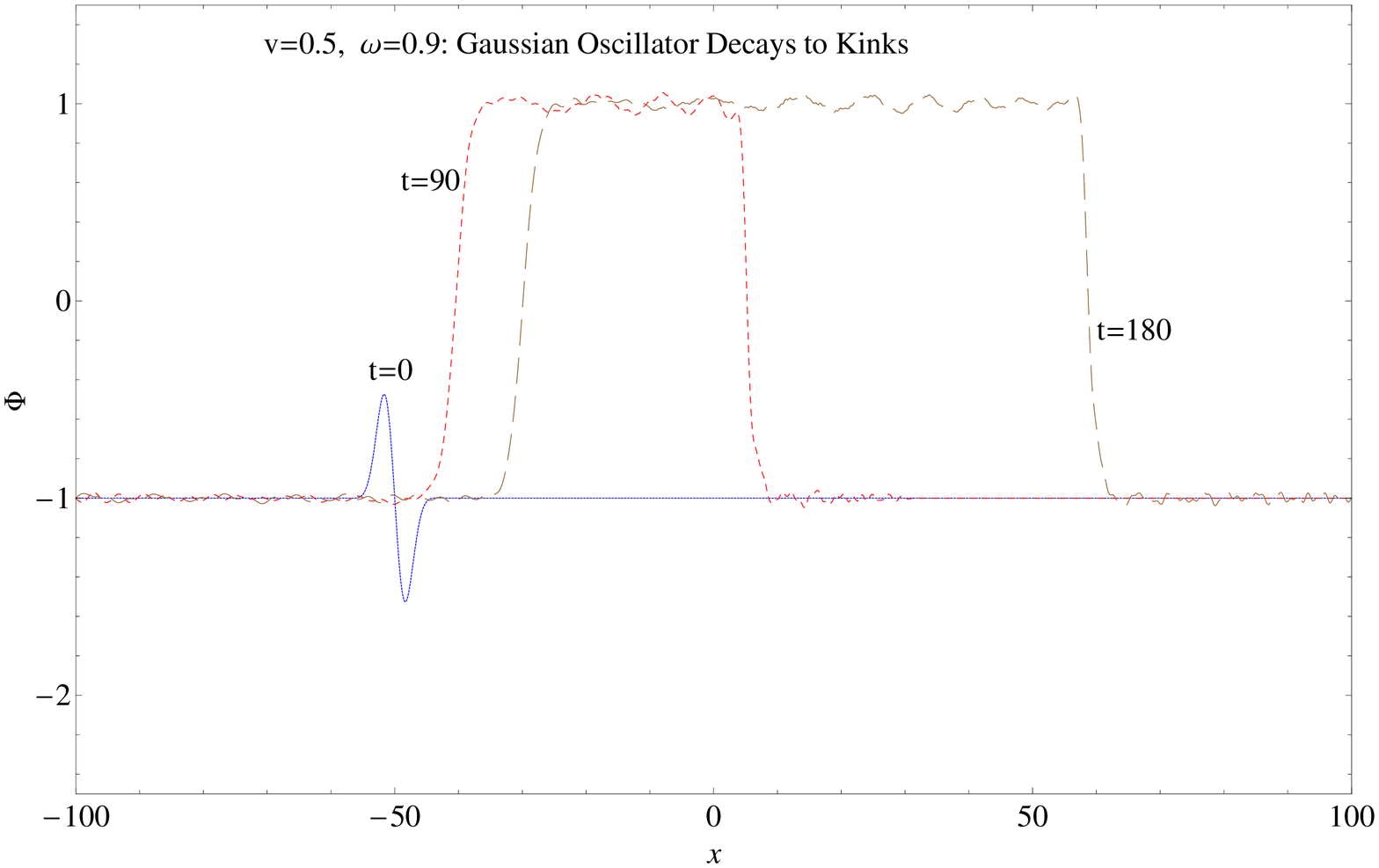} \\
\end{array}$
\end{center}
\vspace{0.0cm}
\caption{\small a: The field evolution of the boosted gaussian oscillator with $(v,\omega)=(0.7,0.5)$, embedded in the $\Phi^4$ Lagrangian. The `particle' evolves oscillating without decay to solitons since there is not enough energy for the decay. b: The field evolution of the boosted gaussian oscillator with $(v,\omega)=(0.5,0.9)$, embedded in the $\Phi^4$ Lagrangian. In this case, the boost provides sufficient energy for the decay to a kink-antikink pair.}
\label{fig6}
\end{figure*}

In order to confirm the above expectations for boosted oscillator decay we have performed numerical simulations for the evolution of the boosted oscillator of equation (\ref{gosc}) in a $\Phi^4$ potential for various $v-\omega$ parameter values. We have confirmed the existence of parameter regions where the decay to a kink-antikink pair occurs and we have found that this region of parameters is in good agreement with the expectations based on energetic arguments. In Fig. 6 we show two types of boosted oscillator evolution corresponding to stability $(v,\omega)=(0.7,0.5)$ and to decay $(v,\omega)=(0.5,0.9)$. The full range of parameters where we observed decay of the oscillator to a kink-antikink pair is spanned by small dots in Fig. 5 while thick dots correspond to the observation of more complicated decay products involving either two kink-antikink pairs or a kink-antikink pair plus particle like oscillators in either of the two vacua. Clearly, the region spanned by the dots where decay occurred in the simulations is in good agreement with the shaded region predicting the decay on the basis of energetic arguments.

\section{Space Dependent Distortions of the Potential}
We now consider numerical experiments in 1+1 dimensions where the dynamical evolution potential is of the following form
\begin{figure}[!t]
\hspace{0pt}\rotatebox{0}{\resizebox{.5\textwidth}{!}{\includegraphics{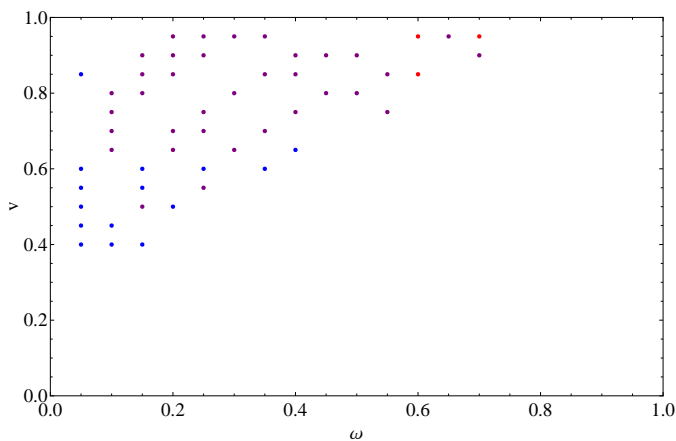}}}
\vspace{0pt}{\caption{The dots indicate parameter values where the decay of the breather into a kink-antikink pair was observed in the simulations at $x_0=0$ where the potential switches to the $\Phi^4$ form. Dots indicate kink-antikink formation from the decay of breather initial conditions, with the different colors indicating the different potentials at which the decay takes place. Blue dots indicate decay in the region of sine-Gordon potential, red dots indicate decay in the region of the $\Phi^4$ potential, and purple dots indicate decay in both regions.}} \label{fig7}
\end{figure}
\ba
V(\Phi)&=&\frac{1}{\pi^2} [ 1+\cos (\pi \Phi)] \;\; x<0 \label{pota}\\
V(\Phi)&=&\frac{1}{4} \left ( \Phi^2 - 1 \right )^2 \; \; x\geq 0 \label{potb}
\ea
ie sine-Gordon for $x<0$ and $\Phi^4$ for $x\geq 0$. Thus, the potential switch occurs now in space rather than in time, at the point $x_0=0$. Our initial condition consists of an oscillating mode (sine-Gordon breather eq. (\ref{br-sol})) with frequency $\omega$ centered at $x=-50$ at $t=0$ boosted with velocity $v$. The boundary conditions are fixed as in the previous section. The results of our numerical experiments are shown in Fig. 7 where the parameter space $v,\omega$ has been spanned at steps of $0.05$.
\begin{figure*}[ht]
\centering
\begin{center}
$\begin{array}{@{\hspace{-0.10in}}c@{\hspace{0.0in}}c}
\multicolumn{1}{l}{\mbox{}} &
\multicolumn{1}{l}{\mbox{}} \\ [-0.2in]
\epsfxsize=3.3in
\epsffile{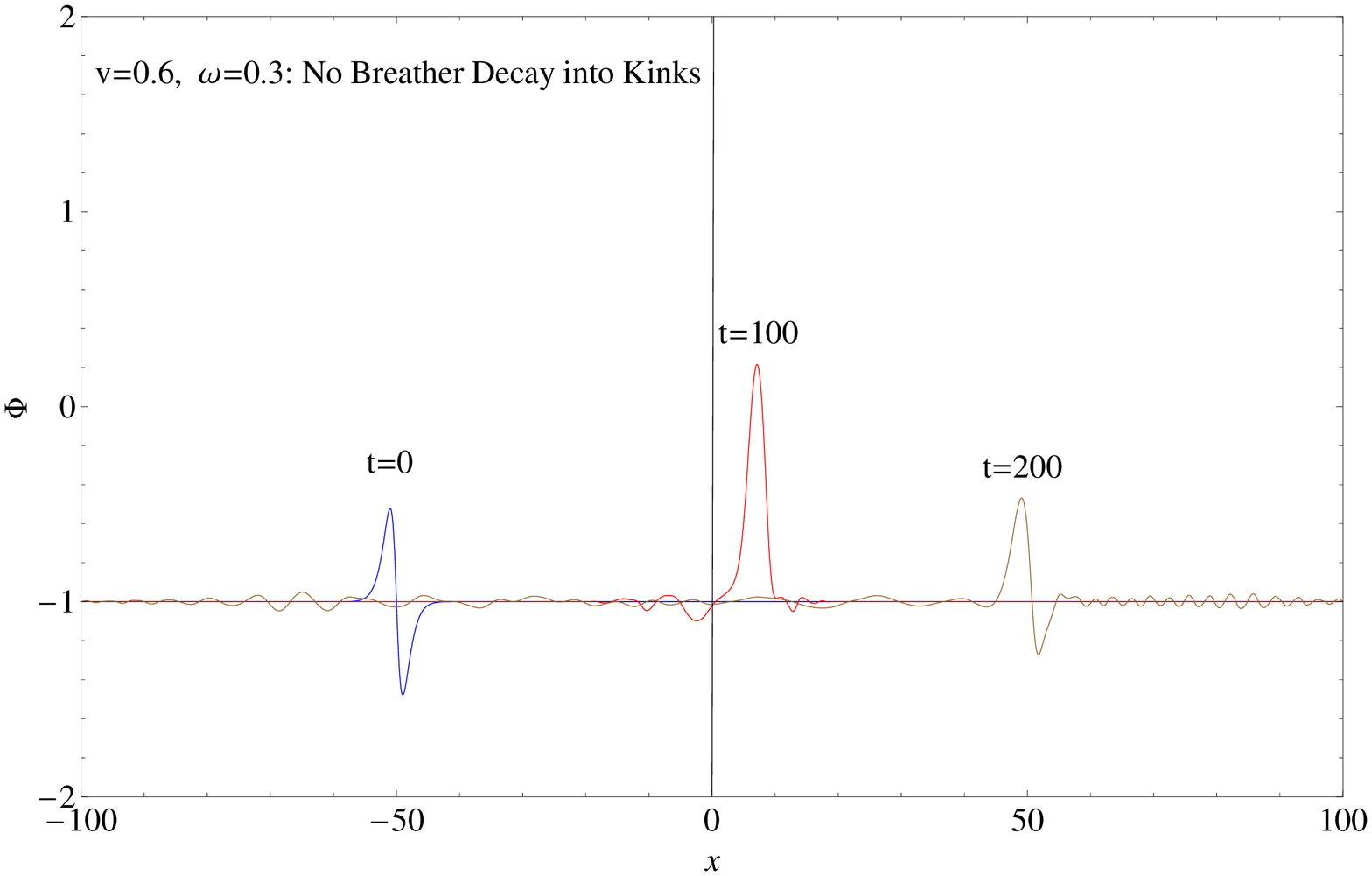} &
\epsfxsize=3.3in
\epsffile{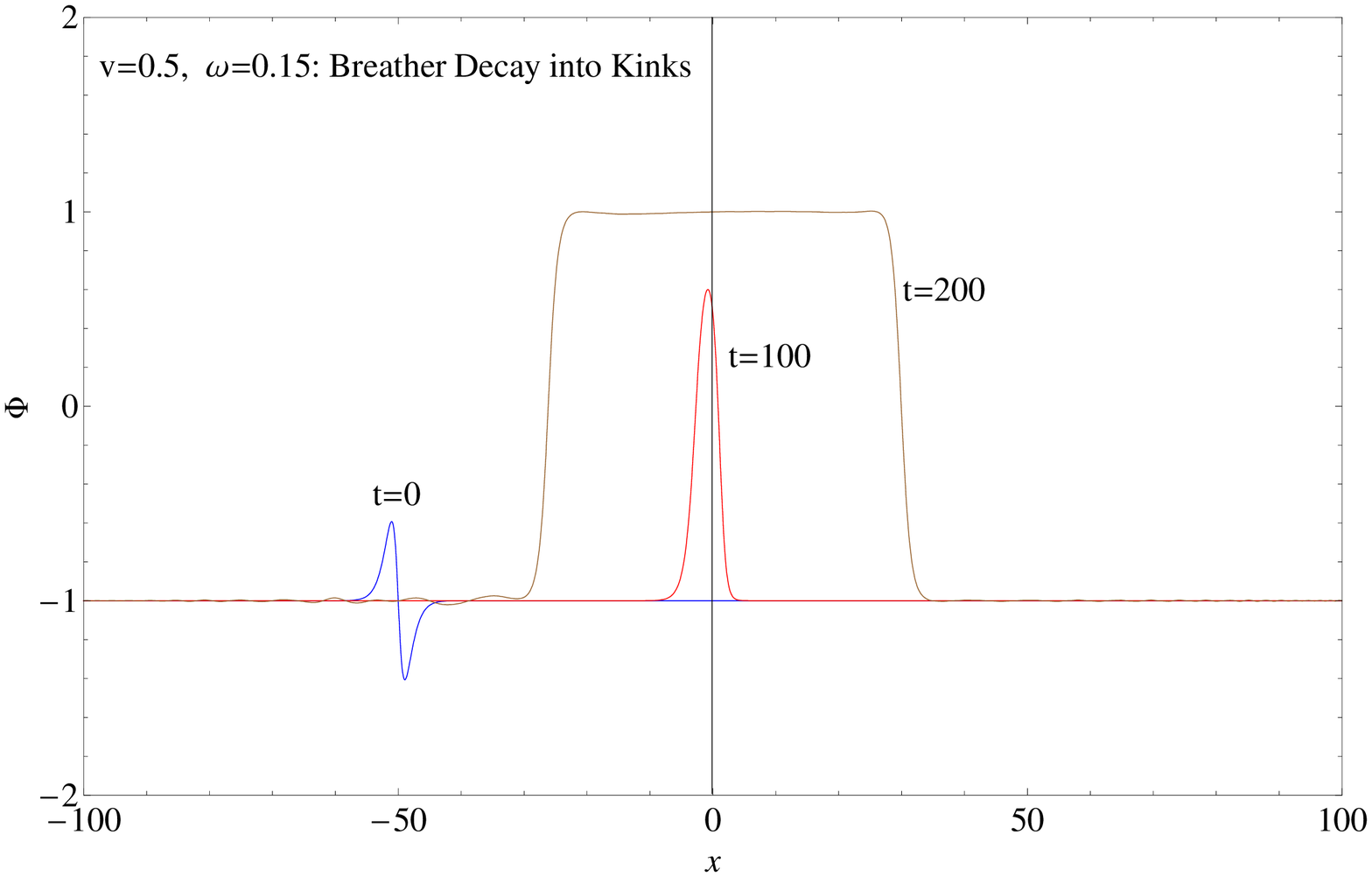} \\
\end{array}$
\end{center}
\vspace{0.0cm}
\caption{\small a: The field evolution of the boosted breather with $(v,\omega)=(0.6,0.3)$, in a potential that switches from sine-Gordon (left) to $\Phi^4$ (right) at $x_0=0$. The `particle' evolves oscillating without decay to solitons for these parameter values. b: The field evolution of the boosted breather with $(v,\omega)=(0.5,0.15)$, in a potential that switches from sine-Gordon (left) to $\Phi^4$ (right) at $x_0=0$. The `particle' evolves oscillating but decays to a kink-antikink pair for these parameter values.}
\label{fig8}
\end{figure*}
These results indicate the following:
\begin{itemize}
\item There is a range of $v,\omega$ parameters (empty regions in Fig. 7) where the oscillator mode is not converted to a kink-antikink pair. Instead, the mode either continues oscillating on the $\Phi^4$ side ($x\geq 0$) or breaks in two modes evolving on both sides of the transition point $x_0=0$.
\item There is a range of $v,\omega$ parameters where the oscillator mode is converted to a kink-antikink pair with the two solitons evolving on both sides of the transition point $x_0=0$ (purple dots in Fig. 7). Alternatively, the two solitons may be reflected back in the sine-Gordon side ($x<0$, blue dots in Fig. 7) or evolve in the $\Phi^4$ side ($x\geq 0$, red dots in Fig. 7).
\end{itemize}
Representative frames from  the simulations corresponding to the above types of evolutions are shown in Fig. 8
\begin{figure}[!t]
\hspace{0pt}\rotatebox{0}{\resizebox{.5\textwidth}{!}{\includegraphics{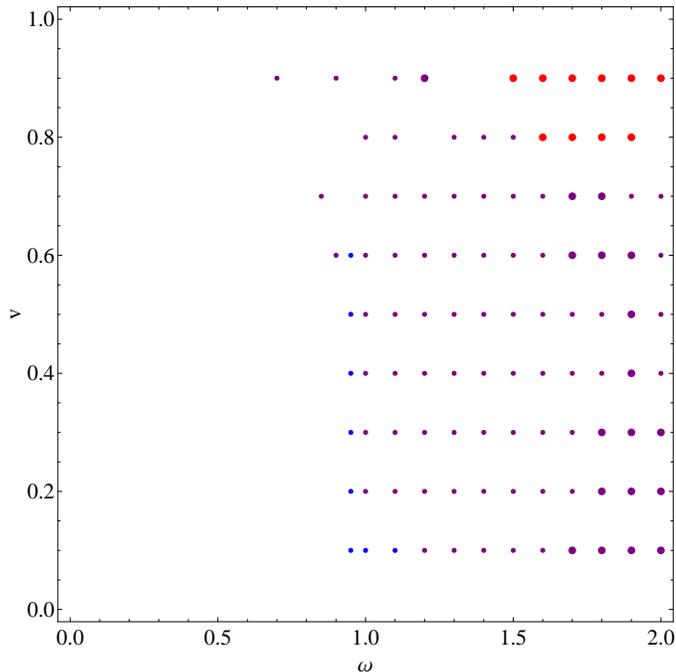}}}
\vspace{0pt}{\caption{The dots indicate parameter values where the decay of the gaussian oscillator into a kink-antikink pair was observed in the simulations at $x_0=0$ where the potential switches to the $\Phi^4$ form. Dots indicate kink-antikink formation, with the different colours indicating the different potentials at which the decay takes place. Blue dots indicate decay in the region of sine-Gordon potential, red dots indicate decay in the region of the $\Phi^4$ potential, and purple dots indicate decay in both regions. Thick dots indicate formation of more complex products. The cases where at least a kink-antikink pair is formed are indicated by a thick dot of the colour corresponding to the potential where the decay occurs, whereas the cases that only new oscillons are formed are indicated by pink think dots. }} \label{fig9}
\end{figure}
Similar results are obtained when using a gaussian oscillator mode as an initial condition. These are shown in Fig. 9

\section{Conclusion-Discussion}
We have shown using numerical simulations that abrupt distortions of the evolution potential in time or in space can lead to a conversion of an oscillation mode to a kink-antikink pair in 1+1 dimensions. The parameter range where this conversion occurs can be approximately obtained in the case of time-dependent distortions by using energetic considerations. We have made no attempt to use such considerations in the case of space dependent distortions due to the complexity of the various possible outcomes of the numerical experiments involving multiple soliton evolution in more complex potential forms.

Since, the oscillating modes are used to represent particles at the classical level, the above described phenomenon may correspond to a new mechanism for soliton formation from particle decay. It may be effective with proper initial conditions in laboratory setups of condensed matter systems as well as high energy systems. The required setup could consist of a high energy particle entering a region of space where the presence of a strong external field would properly modify the field potential responsible for the dynamics. Alternatively, an abrupt change of an external parameter of a system (eg pressure) may lead to abrupt modification of the effective potential determining dynamics. Such experimental setups could reproduce the conditions considered in our numerical simulations leading to soliton formation. In addition this soliton formation mechanism may be relevant in the early universe leading to topological defect formation at temperatures below the phase transition in regions of space where strong fields or pressure gradients are present.

It is straightforward to extend our results to higher dimensional systems. For example oscillation modes in systems accepting vortex solutions may also decay to vortex-antivortex pairs for proper range of parameters $v-\omega$ in the context of a similar mechanism as the one described here. In fact, the reverse process of a vortex-antivortex annihilation to an oscillon was observed in Ref. \cite{Gleiser:2007te}. The investigation of this type of oscillation mode decay to higher dimensional solitons is an interesting subject for future investigation.

The Mathematica file used for the production of the figures may be downloaded from http://leandros.physics.uoi.gr/partkinks.zip .

\section*{Acknowledgements}
We thank T. Vachaspati, D. Steer and S. Nesseris for a careful reading of the paper which lead to insightful comments. C.S.C. thanks the Department of Physics at the University of Ioannina for hospitality during the development of this project. This work was supported by the European Research and Training Network MRTPN-CT-2006 035863-1 (UniverseNet).


\begin{thebibliography}{99}
\bibitem{Rajaramanbook}
``Solitons and Instantons'',
R. Rajaraman, North-Holland, Amsterdam (1987).

\bibitem{Vachaspatibook}
``Kinks and Domain Walls'',
T. Vachaspati, Cambridge University Press (2006).

%\cite{Tong:2005un}
\bibitem{Tong:2005un}
  D.~Tong,
  %``TASI lectures on solitons,''
  arXiv:hep-th/0509216.
  %%CITATION = HEP-TH/0509216;%%

%\cite{Dutta:2008jt}
\bibitem{Dutta:2008jt}
  S.~Dutta, D.~A.~Steer and T.~Vachaspati,
  %``Creating Kinks from Particles,''
  Phys.\ Rev.\ Lett.\  {\bf 101}, 121601 (2008)
  [arXiv:0803.0670 [hep-th]].
  %%CITATION = PRLTA,101,121601;%%


\bibitem{Bunkov:2000} Y.M. Bunkov, H. Godfin (Eds.), `Topological Defects and the Non-Equilibrium
Dynamics of Symmetry Breaking Phase Transitions', Kluwer Academic Publ.,
Dordrecht/Boston/London 2000.

\bibitem{Arodz:2003} H. ArodzŸ , J. Dziarmaga, W.H. Zurek (Eds.), `Patterns of Symmetry Breaking',
Kluwer Academic Publ., Dordrecht/Boston/London 2003.


\bibitem{def-condmat}
  K.~J.~M.~Moriarty, E.~Myers and C.~Rebbi,
  %``Dynamical Interactions Of Flux Vortices In Superconductors,''
  Phys.\ Lett.\  B {\bf 207}, 411 (1988); M.~Donaire, T.~W.~B.~Kibble and A.~Rajantie,
  %``Spontaneous vortex formation on a superconductor film,''
  New J.\ Phys.\  {\bf 9}, 148 (2007)
  [arXiv:cond-mat/0409172]; W.~H.~Zurek,
  %``Cosmological Experiments in Condensed Matter Systems,''
  Phys.\ Rept.\  {\bf 276}, 177 (1996)
  [arXiv:cond-mat/9607135]; V.~M.~Ruutu, V.~B.~Eltsov, M.~Krusius, Yu.~G.~Makhlin, B.~Placais and G.~E.~Volovik,
  %``Defect Formation in Quench-Cooled Superfluid Phase Transition,''
  Phys.\ Rev.\ Lett.\  {\bf 80}, 1465 (1998);
  T.~W.~B.~Kibble,
  %``Testing cosmological defect formation in the laboratory,''
  arXiv:cond-mat/0111082; I.~Chuang, B.~Yurke, A.~N.~Pargellis and N.~Turok,
  %``Coarsening dynamics in uniaxial nematic liquid crystals,''
  Phys.\ Rev.\  E {\bf 47}, 3343 (1993).
  %%CITATION = PHRVA,E47,3343;%%


%\cite{Hindmarsh:1994re}
\bibitem{Hindmarsh:1994re}
  M.~B.~Hindmarsh and T.~W.~B.~Kibble,
  %``Cosmic strings,''
  Rept.\ Prog.\ Phys.\  {\bf 58}, 477 (1995)
  [arXiv:hep-ph/9411342].
  %%CITATION = RPPHA,58,477;%%

%\cite{Durrer:2001cg}
\bibitem{Durrer:2001cg}
  R.~Durrer, M.~Kunz and A.~Melchiorri,
  %``Cosmic structure formation with topological defects,''
  Phys.\ Rept.\  {\bf 364}, 1 (2002)
  [arXiv:astro-ph/0110348].
  %%CITATION = PRPLC,364,1;%%

%\cite{Perivolaropoulos:2005wa}
\bibitem{Perivolaropoulos:2005wa}
  L.~Perivolaropoulos,
  %``The rise and fall of the cosmic string theory for cosmological
  %perturbations,''
  Nucl.\ Phys.\ Proc.\ Suppl.\  {\bf 148}, 128 (2005)
  [arXiv:astro-ph/0501590].
  %%CITATION = NUPHZ,148,128;%%

%\cite{Magueijo:2000se}
\bibitem{Magueijo:2000se}
  J.~Magueijo and R.~H.~Brandenberger,
  %``Cosmic defects and cosmology,''
  arXiv:astro-ph/0002030.
  %%CITATION = ASTRO-PH/0002030;%%


%\cite{Levkov:2004tf}
\bibitem{Levkov:2004tf}
  D.~G.~Levkov and S.~M.~Sibiryakov,
  %``Induced tunneling in QFT: Soliton creation in collisions of highly
  %energetic particles,''
  Phys.\ Rev.\  D {\bf 71}, 025001 (2005)
  [arXiv:hep-th/0410198].
  %%CITATION = PHRVA,D71,025001;%%

%\cite{Mattis:1991bj}
\bibitem{Mattis:1991bj}
  M.~P.~Mattis,
  %``The Riddle of high-energy baryon number violation,''
  Phys.\ Rept.\  {\bf 214}, 159 (1992).
  %%CITATION = PRPLC,214,159;%%

%\cite{Manton:1996ex}
\bibitem{Manton:1996ex}
  N.~S.~Manton and H.~Merabet,
  %``phi~4 Kinks - Gradient Flow and Dynamics,''
  arXiv:hep-th/9605038.
  %%CITATION = HEP-TH/9605038;%%

%\cite{Romanczukiewicz:2005rm}
\bibitem{Romanczukiewicz:2005rm}
  T.~Romanczukiewicz,
  %``Creation of kink and antikink pairs forced by radiation,''
  J.\ Phys.\ A  {\bf 39}, 3479 (2006)
  [arXiv:hep-th/0501066].
  %%CITATION = JPAGB,A39,3479;%%

\bibitem{Colemanbook}
``Aspects of Symmetry: Selected Erice Lectures'',
S. Coleman, Cambridge University Press (1985).

%\cite{Gleiser:2007te}
\bibitem{Gleiser:2007te}
  M.~Gleiser and J.~Thorarinson,
  %``A phase transition in U(1) configuration space: Oscillons as remnants of
  %vortex antivortex annihilation,''
  arXiv:hep-th/0701294.
  %%CITATION = HEP-TH/0701294;%%

%\cite{Gleiser:1993pt}
\bibitem{Gleiser:1993pt}
  M.~Gleiser,
  %``Pseudostable bubbles,''
  Phys.\ Rev.\  D {\bf 49}, 2978 (1994)
  [arXiv:hep-ph/9308279].
  %%CITATION = PHRVA,D49,2978;%%

\bibitem{rindler88}W. Rindler, J. Denur, Am. J. Phys., {\bf 56}, (9), 795 (1988).

\bibitem{mathematica} http://www.wolfram.com/

%\cite{Bogolyubsky:1976nx}
\bibitem{Bogolyubsky:1976nx}
  I.~L.~Bogolyubsky and V.~G.~Makhankov,
  %``On The Pulsed Soliton Lifetime In Two Classical Relativistic Theory
  %Models,''
  JETP Lett.\  {\bf 24}, 12 (1976).
  %%CITATION = JTPLA,24,12;%%

%\cite{Copeland:1995fq}
\bibitem{Copeland:1995fq}
  E.~J.~Copeland, M.~Gleiser and H.~R.~Muller,
  %``Oscillons: Resonant configurations during bubble collapse,''
  Phys.\ Rev.\  D {\bf 52}, 1920 (1995)
  [arXiv:hep-ph/9503217].
  %%CITATION = PHRVA,D52,1920;%%

%\cite{Honda:2000gv}
\bibitem{Honda:2000gv}
  E.~P.~Honda,
  %``Resonant dynamics within the nonlinear Klein-Gordon equation: Much ado
  %about oscillons,''
  arXiv:hep-ph/0009104.
  %%CITATION = HEP-PH/0009104;%%

%\cite{Hindmarsh:2006ur}
\bibitem{Hindmarsh:2006ur}
  M.~Hindmarsh and P.~Salmi,
  %``Numerical investigations of oscillons in 2 dimensions,''
  Phys.\ Rev.\  D {\bf 74}, 105005 (2006)
  [arXiv:hep-th/0606016].
  %%CITATION = PHRVA,D74,105005;%%

\end{thebibliography}
\end{document}